%% file: document.tex
\documentclass[prl,superscriptaddress,twocolumn,showpacs,preprintnumbers,amsmath,amssymb]{revtex4}

\usepackage{epsfig}
\usepackage{dcolumn}
\usepackage{bm}

\usepackage{graphicx}

\include{include}

\begin{document}

\title{ \boldmath Observation of $WZ$ Production}

\input{authorList/authorList.tex}

\begin{abstract} 

We report the first observation of the associated production of a $\W$
boson and a $\Z$ boson. This result is based on 1.1 ${\rm fb}^{-1}$ of
integrated luminosity from $p\overline{p}$ collisions at $\sqrt{s} =
1.96~{\rm TeV}$ collected with the CDF II detector at the Fermilab
Tevatron. We observe 16 $\WZ$ candidates passing our event selection
with an expected background of $2.7 \pm 0.4$ events. A fit to the
missing transverse energy distribution indicates an excess of events
compared to the background expectation corresponding to a significance
equivalent to six standard deviations. The measured cross section is 
$\sigma(p\overline{p}\rightarrow WZ) = 5.0^{+1.8}_{-1.6}$ pb,
consistent with the standard model expectation.

\end{abstract}

\pacs{12.15.Ji 13.40.Em 13.87.Ce 14.70.Fm 14.70.Hp}

\maketitle

\normalsize


The $\W$ and $\Z$ vector bosons which mediate the weak interaction
are produced individually in large numbers in $\ppbar$ collisions at 
the Fermilab Tevatron, and their cross sections have been measured 
with high precision \cite{Acosta:2004uq}. The production of heavy vector
boson pairs ($\WW$, $\WZ$, and $\ZZ$) is far less common and can
involve the triple gauge couplings (TGCs) between the bosons themselves 
via an intermediate virtual boson. Deviations of measured diboson
production properties from standard model (SM) predictions could arise
from new interactions or loop effects due to new particles at energy
scales not directly accessible to a given experiment
\cite{PhysRevD.48.2182}. At the Tevatron, TGCs are probed at the
highest energy scales yet achieved.

In this Letter, we report the first observation of $WZ$
production. The production is observed in $p\overline{p}$ collisions
at $\sqrt{s} = 1.96~{\rm TeV}$ using 1.1 ${\rm fb}^{-1}$ of integrated
luminosity collected by the CDF II detector at the Fermilab
Tevatron. We consider the decay channel $\WZdecay$, where
$\ell^{\prime}$ and $\ell$ are electrons or muons directly from $\W$
and $\Z$ decay, respectively, or from the leptonic decay of $\tau$'s
when one or both vector bosons decay to $\tau$ leptons.

The most sensitive previous search for $\WZ$ production was reported
by the $\Dzero$ Collaboration using 0.3 ${\rm fb}^{-1}$ of integrated
luminosity, where three $\WZdecay$ candidate events were found
\cite{Abazov:2005ys}. The observed events had a probability of 3.5\%
to be due to background fluctuations, corresponding to $\sigma(WZ) <
13.3$ pb at 95\% C.L. A search for the sum of $\WZ$ and $\ZZ$
production in decays to 2, 3, and 4 lepton channels by the CDF
Collaboration using 0.194 ${\rm fb}^{-1}$ of integrated luminosity
determined that $\sigma(WZ+ZZ) < 15.2$ pb at 95\%
C.L. \cite{Acosta:2005pq}. The next-to-leading order (NLO) $\WZ$ cross
section prediction for $p\overline{p}$ collisions at $\sqrt{s} =
1.96~{\rm TeV}$ is 3.7 $\pm$ 0.3 pb \cite{Campbell:1999ah}.


The components of the CDF II detector relevant to this analysis are
described briefly here; a more complete description can be found
elsewhere \cite{Acosta:2004yw}. The detector geometry is described using
the azimuthal angle $\phi$ and the pseudorapidity $\eta
\equiv-\ln[\tan(\theta/2)]$, where $\theta$ is the polar angle with
respect to the proton beam axis (positive $z$-axis). The
pseudorapidity of a particle originating from the center of the
detector is referred to as $\eta_{\rm{d}}$.

The trajectories of charged particles (tracks) are reconstructed using
silicon microstrip detectors \cite{SVXII,ISL} and a 96-layer open-cell
drift chamber (COT) \cite{COT} inside a 1.4~T solenoid. The number of
COT layers traversed by a particle in the range $\etad \leq 1$ is 96
and decreases to zero for $\etad\rightarrow 2$. The silicon system
provides coverage with 6 (7) layers with radii between 2.4 cm and 28
cm for $\etad<1.0~(1.0<\etad<2.0)$. Outside of the solenoid are
electromagnetic (EM) and hadronic (HAD) sampling calorimeters,
segmented in a projective tower geometry, and constructed of layers of
lead or iron absorber, respectively, and scintillator. The EM section
is the first 19-21 radiation lengths ($X_0$), corresponding to one hadronic 
interaction length ($\lambda$) and contains electromagnetic showers, while 
the HAD section extends to 4.5-7 $\lambda$ and contains the majority of a 
hadronic shower. The calorimeters are divided into central
($\etad<1.1$) and forward ($1.1<\etad<3.64$) regions. Outside of the
central calorimeters are muon detectors consisting of scintillators
and drift chambers.


Including the leptonic $\tau$ decays, the branching fraction of the
$\WZ$ state to three $e$ or $\mu$ leptons is 1.8\%. When coupled
with the small SM cross section, this implies that only a small number
($\sim$70) of $\WZdecay$ events are expected to be produced in 1.1 ${\rm
fb}^{-1}$ at the Tevatron. Furthermore, in order to identify a $\WZ$
event in this decay channel, all three charged leptons must be
detected. The CDF II detector, however, has gaps in calorimeter
coverage and limited forward ($\etad \simge 1$) tracking
efficiency.

In order to maximize the total acceptance, while minimizing the
backgrounds from jets and photons misidentified as leptons, we exploit
all available reconstructed tracks and clusters of energy in the
EM calorimeter. We separate these into seven non-overlapping lepton
categories: three each of electrons and muons, and a seventh for
tracks that are non-fiducial to the calorimeters and are not
identified as muons. In this context, ``non-fiducial'' refers to
detector regions that are inactive for energy measurement because they
are either not covered, or are only partially covered, by calorimeter
components.

All lepton candidates are required to be isolated such that the sum
of the $E_{T}$ for the calorimeter towers in a cone of $\Delta R =
\sqrt{(\Delta\eta)^2 + (\Delta\phi)^2} < 0.4$ around the lepton is
less than 10\% of the $\Et$ for electrons or $\pt$ for muons and track
lepton candidates. The transverse energy $\Et$ of an energy cluster or
calorimeter tower is $E\sin\theta$, where $E$ is the associated
energy. Similarly, $p_{T}$ is the component of track momentum
transverse to the beam line.

All electron candidates are required to have a cluster of energy in
the calorimeter with the ratio of deposition in the HAD to EM sections
consistent with being due to an electron. These candidates are divided
into three categories: those in the central calorimeter, those in the
forward calorimeter matched to a track, and those in the forward
calorimeter without a matched track. The central electron category
requires a well-measured COT track. Since the tracking efficiency is
low in the large $\etad$ region, a track pattern algorithm which
starts with calorimeter information and attempts to attach silicon
hits is used for forward electrons. For forward electrons without a
matched track, both charge hypotheses are considered when forming $WZ$
candidates, since the charge is determined from the track curvature.

For all muon candidates, the energy deposition in both the EM and HAD
calorimeter sections is required to be consistent with that of a minimum
ionizing particle. The muon candidates are divided into a category in
which the tracks match to reconstructed track segments (``stubs'') in
the muon chambers and two categories of tracks that do not match to
stubs (``stubless''). The stubless muon candidates are designated as
central or forward, depending on the calorimeter to which the track is
projected. The stubbed and central stubless muons have strict
requirements on the number of COT hits and the $\chi^2$ of the track
fit in order to suppress background muons from $K^\pm$ or $\pi^\pm$
decays. To increase the track finding efficiency in the forward
region, we use an algorithm that starts with silicon detector hits in
addition to one that starts with COT hits. The forward stubless muons
require at least 60\% of the traversed COT layers to have hits. To
suppress the background from cosmic rays and $K^\pm$ or $\pi^\pm$
decays, we require the point of closest approach of the track to the
beamline to be consistent with having originated from the beam, in
addition to using a cosmic ray rejection algorithm.

An additional category consists of tracks that neither project
to the fiducial regions of the calorimeters nor are identified as
stubbed muons. The requirements for these track-only lepton candidates
are the same as for central stubless muons, but without the
calorimeter requirements. Due to the lack of calorimeter information,
electrons and muons cannot be reliably differentiated for this category,
and are therefore treated as having either flavor in the $\WZ$
candidate selection. If an electron or track-only candidate is
consistent with a photon conversion, as indicated by the presence of
an additional nearby track with a common vertex, the candidate is
rejected.


To measure the presence of a neutrino, we use missing transverse
energy $\MET = | \sum_{i} E_{T,i} ~ \hat{n}_{T,i} |$, where
$\hat{n}_{T,i}$ is the transverse component of the unit vector
pointing from the interaction point to calorimeter tower $i$. The
$\MET$ calculation is corrected for muons and track-only lepton
candidates, which do not deposit all of their energy in the
calorimeter.


The events we consider must pass one of four online trigger
selections. The events with central electrons require an EM energy
cluster with $\Et >18$ GeV matched to a track with $\pt >
8~\GeVc$. The events with forward electrons require an EM energy
cluster with $\Et >20$ GeV and an uncorrected, calorimeter-based
measurement of $\MET > 15$ GeV. Muon triggers are based on stubs from
the muon chambers matched to a track with $\pt > 18~\GeVc$. Trigger
efficiencies are measured in leptonic $W$ and $Z$ data samples
\cite{Acosta:2004uq}.


The $WZ$ candidates are selected from events with exactly three lepton
candidates using requirements that were optimized with Monte Carlo
simulation without reference to the data. At least one lepton is
required to satisfy the trigger and have $\Et > 20$ GeV ($\pt >
20~\GeVc$) for electrons (muons). We loosen this requirement to $10$
GeV ($\GeVc$) for the other leptons to increase the $WZ$ kinematic
acceptance. Aside from $WZ$ production, other SM processes that can
lead to three high-$\pt$ leptons include dileptons from the Drell-Yan
$Z/\gamma^{*}$ process (DY), with an additional lepton from a photon
conversion ($Z\gamma$) or a misidentified jet ($Z$+jets) in the event;
$ZZ$ production where only three leptons are identified and the
unobserved lepton results in $\MET$; and a small contribution from
$\ttbar \rightarrow \W b\W \bar{b}$, where two charged leptons result
from the $W$ boson decays and one or more from decay of the
$b$-quarks. Except for $\ttbar$, these backgrounds are suppressed by
requiring $\MET>25$ GeV in the event, consistent with the unobserved
neutrino from the leptonic decay of a $W$ boson. We also require the
azimuthal angle between the $\MET$ direction and any identified jet
with $\Et > 15$ GeV or $WZ$ candidate lepton to be greater than
$9^\circ$ to suppress DY backgrounds in which the observed $\MET$ is
due to mismeasured leptons and/or jets.

We require at least one same-flavor, opposite-sign lepton pair in the
event with an invariant mass $M_{\ell^{+}\ell^{-}}$ in the range [76,
106] $\GeV$, consistent with the decay of a $Z$ boson. This range 
is referred to as the ``$\Z$-mass region.'' If there is more than one
such pair, the leptons with $M_{\ell^{+}\ell^{-}}$ closest to the $Z$
mass \cite{PDG} are treated as the $Z$ boson decay candidate pair. In
order to suppress backgrounds from $ZZ$, we require that no additional
track in the event with $\pt>8~\GeVc$, when combined with the lepton
that is not part of the $Z$ boson decay candidate pair, has an
invariant mass in the $Z$-mass region. The overall acceptance for
$\WZdecay$, using the described selection criteria, is 13.4\%. 


The acceptances for the $\WZ$, $\ZZ$, $Z\gamma$, and $\ttbar$
processes are determined using Monte Carlo calculations followed by a
{\sc geant}-based simulation \cite{GEANT} of the CDF II detector. The
Monte Carlo generator used for $\WZ$, $\ZZ$, and $\ttbar$ is {\sc
pythia} \cite{Sjostrand:2006za} and for $Z\gamma$ is the generator
described in \cite{Baur:1992cd}. We use the {\sc CTEQ5L} parton
distribution functions (PDFs) to model the longitudinal momentum
distribution of the initial-state partons \cite{Lai:1999wy}. An
efficiency correction, of up to 10\% per lepton, is applied to the
simulation based on measurements of the lepton reconstruction and
identification efficiencies using observed
$\Z\rightarrow\ell^{+}\ell^{-}$ events. An additional correction is
applied to the $Z\gamma$ background estimate based on a measurement of
the photon conversion veto efficiency in data. The background from
$Z$+jets is estimated from a sample of events with two identified
leptons and a jet that is required to pass loose isolation
requirements and contain a track or energy cluster similar to those
required in the lepton identification. The contribution of each event
to the total yield is scaled by the probability that the jet is
identified as a lepton. This probability is determined from multijet
events collected with a set of jet-based triggers. A
correction is applied for the small real lepton contribution using
single $\W$ and $\Z$ boson Monte Carlo simulation.


Systematic uncertainties associated with the Monte Carlo simulation
affect the $Z\gamma$, $ZZ$, $\ttbar$, and $WZ$ simulations similarly.
The uncertainties from the lepton selection and trigger efficiency
measurements are propagated through the analysis, giving 
uncertainties of $1.2\% - 2.0\%$ and $0.4\% - 0.9\%$ for the respective 
efficiencies of the different signal and background processes. 
The uncertainty due to the $\MET$ resolution modeling is determined
from comparisons of the data and the Monte Carlo simulation in a
sample of dilepton events. For $WZ$, $ZZ$, and $\ttbar$ production,
where it is an observed particle that produces the observed $\MET$, we
determined the uncertainty to be $1\%$. The uncertainty due to the
$\MET$ modeling for the $Z\gamma$ background is much larger ($25\%$)
than for other final states because it depends on the non-Gaussian
tails of the resolution function. The uncertainties on the $ZZ$,
$Z\gamma$, and $\ttbar$ cross sections are assigned to be 10\% 
\cite{Campbell:1999ah}, 10\% \cite{Baur:1997kz}, and 15\%
\cite{Kidonakis:2003qe,Cacciari:2003fi}, respectively. For the
$Z\gamma$ background contribution, there is an additional uncertainty
of 20\% from the detector material description and conversion veto
efficiency. The detector acceptance variation due to PDF uncertainties 
is assessed to be $2\%$ using the 20 pairs of PDF sets described in 
\cite{Kretzer:2003it}. The systematic uncertainty on the $Z$+jets
background is determined from differences in the measured probability
that a jet is identified as a lepton for jets collected using
different jet $\Et$ trigger thresholds. These variations correspond to
changing the parton composition of the jets and the relative amount of
contamination from real leptons. The total systematic uncertainty on
the $\Z$+jets background prediction is 23\%. The signal and background
estimates obtained from simulation have an additional 6\% uncertainty
originating from the luminosity measurement \cite{Acosta:2002hx}.


In addition to the signal region ($\MET > 25$ and $\Mll$ in the
$\Z$-mass region), we define two independent trilepton control
regions, both with $\MET < 25$ GeV, but different $\Mll$ criteria, to
validate our background estimates. The ``$Z$-mass control region'' is
defined to have $\Mll$ in the $\Z$-mass region and is dominated by
$\Z$+jets and $\Z\gamma$ where the photon is from initial-state
radiation. The ``$Z$-veto control region'' is defined to have $\Mll$
outside of the $\Z$-mass region and a minimum value for $\Mll$ of
$40~\GeV$. This region is dominated by $\Z\gamma$ where the photon is
from final-state radiation. We expect 243.5 $\pm$ 38.8 (250.9 $\pm$
48.3) events in the $Z$-mass ($Z$-veto) control region and observe 215
(241) events. The results for the trilepton classifications we
consider are shown in Table~\ref{tbl:control_regions} and are in good
agreement with the expectations.

\input{WZ_control_regions}

\begin{figure*}[t]
\unitlength=0.33\linewidth
\includegraphics[width=0.333\linewidth]{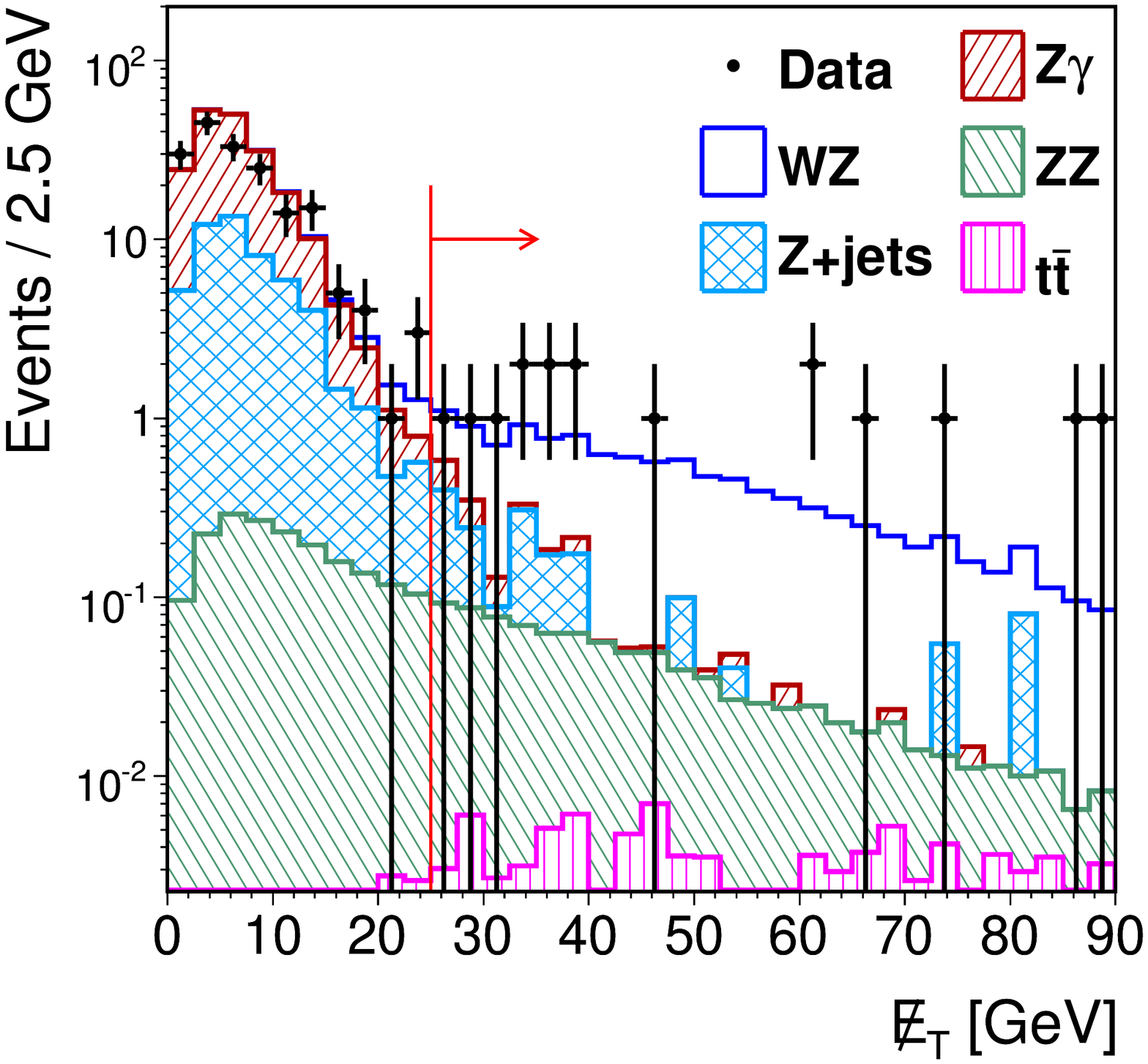}
\put(-0.70,0.80){(a)} 
\includegraphics[width=0.333\linewidth]{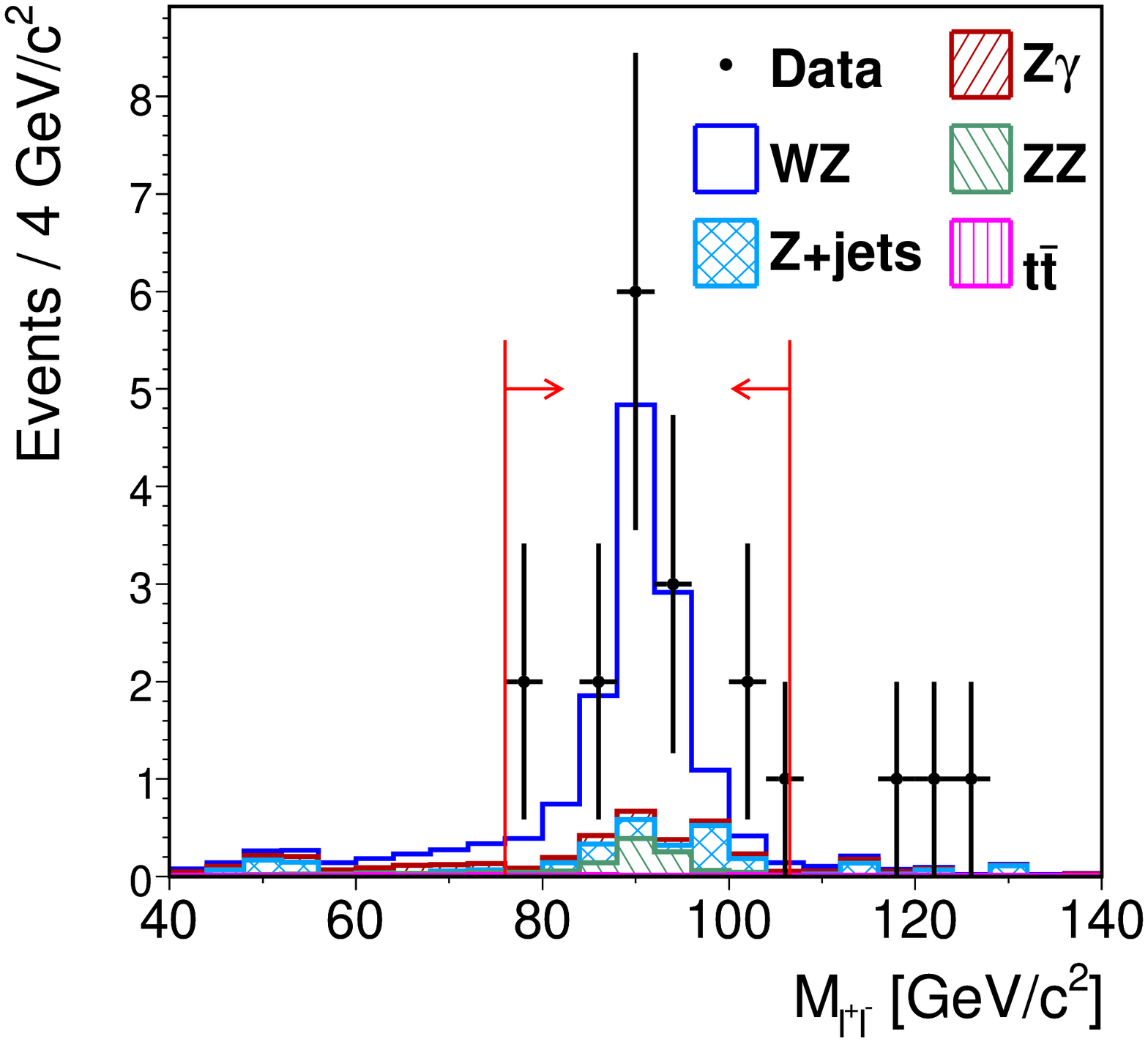}
\put(-0.70,0.80){(b)} 
\includegraphics[width=0.333\linewidth]{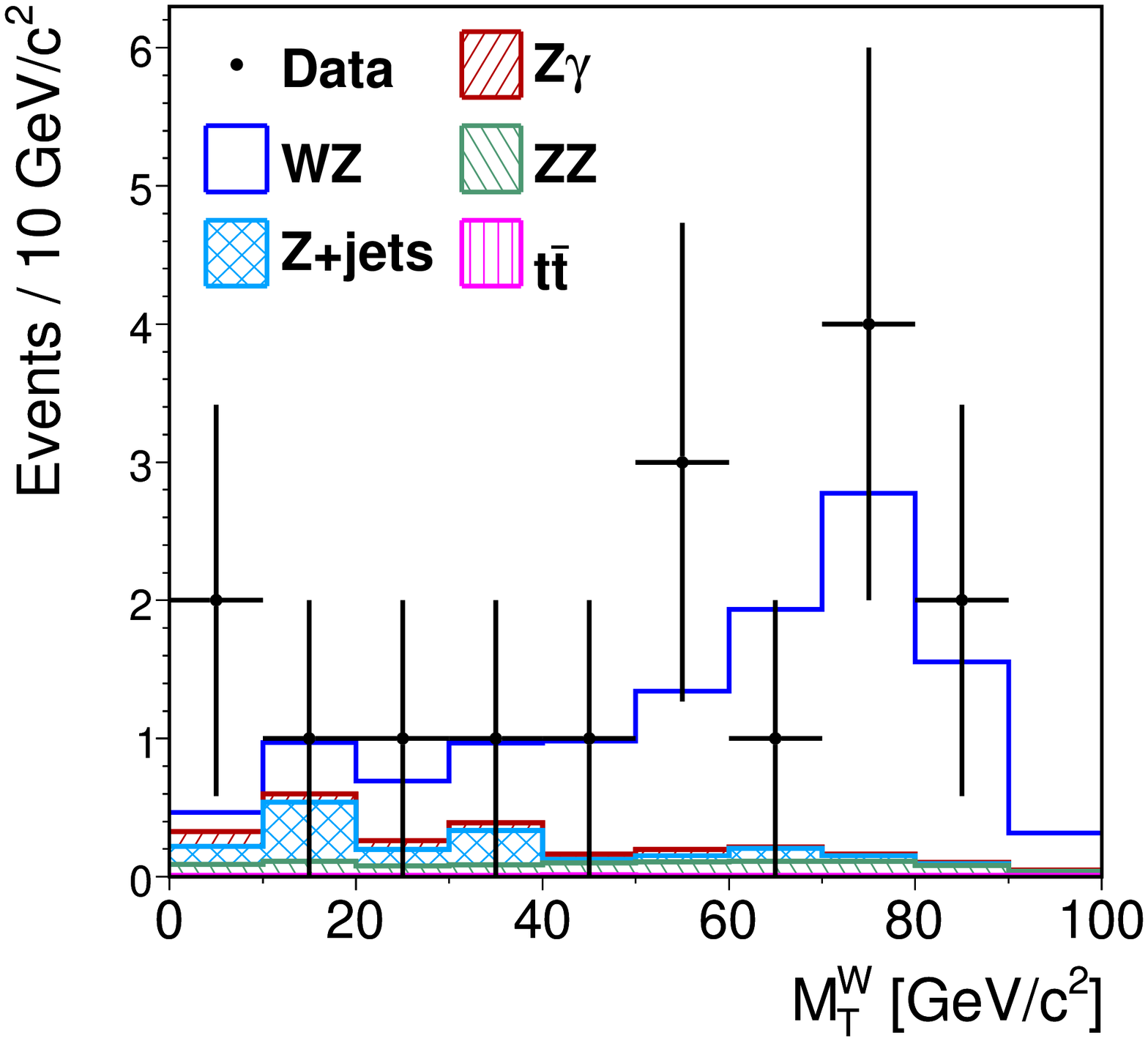}
\put(-0.19,0.80){(c)} 
\caption{Distributions for $WZ$ candidates of (a) the $\MET$ (b) the
dilepton invariant mass for the same-flavor opposite-sign dilepton
pair closest to the $Z$ mass, and (c) the $W$ transverse mass
calculated from the remaining lepton and the $\MET$. In (a) and (b),
the arrows indicate the signal region.} 
\label{fig:kinematics}
\end{figure*}

The expectation in the signal region for $WZ$ and each background
contribution is summarized in Table~\ref{tbl:results}. In this region,
we expect 2.7 $\pm$ 0.4 background events and observe 16 events. A
breakdown of the observed (expected) events by flavor classification
is as follows: 6 (2.7 $\pm$ 0.2) $eee$, 0 (2.0 $\pm$ 0.2) $ee\mu$, 1
(1.5 $\pm$ 0.1) $e\mu\mu$, 1 (1.2 $\pm$ 0.1) $\mu\mu\mu$, 5 (2.0 $\pm$
0.2) $ee\ell_t$, 2 (1.3 $\pm$ 0.1) $e\mu \ell_t$, 1 (1.1 $\pm$ 0.1)
$\mu\mu \ell_t$, 0 (0.5 $\pm$ 0.1) $e\ell_t\ell_t$, and 0 (0.2 $\pm$
0.1) $\mu \ell_t\ell_t$. Here $\ell_t$ denotes the track-only lepton
candidates having unknown flavor. The distributions of $\MET$, $\Mll$,
and the $W$ transverse mass $M_{T}^{W} \equiv \sqrt{2 E_{T} \MET (1 -
\cos\phi_{\ell\nu})}$, where $\phi_{\ell\nu}$ is the azimuthal angle
between the non-$Z$ candidate lepton and the $\MET$ direction are
shown in Fig.~\ref{fig:kinematics}. The data are in good agreement
with the SM prediction.

\input{WZ_expect_withsys}

Due to the unobserved neutrino, events from $\WZdecay$ are expected to
have larger $\MET$ on average than the DY and $ZZ$ backgrounds. We
exploit this information by performing a binned maximum likelihood fit
for the signal yield using the following $\MET$ bins: $25 < \MET <
45$ GeV and $\MET > 45$ GeV. This binning was chosen to maximize our
sensitivity for a $\WZ$ observation without reference to the data. We
expect 2.0 $\pm$ 0.4 (0.7 $\pm$ 0.1) background events and 6.5 $\pm$
0.5 (5.9 $\pm$ 0.4) $\WZ$ events, with 9 (7) events observed in the
lower (upper) $\MET$ bin. We define ${\Delta\ln\cal L}$ as the log of
the likelihood ratio between this fit and the no signal
hypothesis. For our data, we find ${2\Delta\ln\cal L}= 37.8$. We
interpret this result using $10^{10}$ background-only Monte Carlo
experiments, out of which only 11 had a larger value of
${2\Delta\ln\cal L}$ (a probability of 1.1$\times10^{-9}$),
corresponding to a significance equivalent to six standard
deviations.

This result represents the first observation of $\WZ$ production. The
measured cross section is
\begin{equation*}
\sigma(p\overline{p}\rightarrow \WZ) = 5.0^{+1.8}_{-1.4}~\rm{(stat.)}\pm~0.4~\rm{(syst.)}~pb,
\end{equation*}
consistent with the SM expectation. 


We thank the Fermilab staff and the technical staffs of the
participating institutions for their vital contributions. This work
was supported by the U.S. Department of Energy and National Science
Foundation; the Italian Istituto Nazionale di Fisica Nucleare; the
Ministry of Education, Culture, Sports, Science and Technology of
Japan; the Natural Sciences and Engineering Research Council of
Canada; the National Science Council of the Republic of China; the
Swiss National Science Foundation; the A.P. Sloan Foundation; the
Bundesministerium f\"ur Bildung und Forschung, Germany; the Korean
Science and Engineering Foundation and the Korean Research Foundation;
the Particle Physics and Astronomy Research Council and the Royal
Society, UK; the Institut National de Physique Nucleaire et Physique
des Particules/CNRS; the Russian Foundation for Basic Research; the
Comisi\'on Interministerial de Ciencia y Tecnolog\'{\i}a, Spain; the
European Community's Human Potential Programme under contract
HPRN-CT-2002-00292; and the Academy of Finland.

\bibliography{document}
\end{document}

%% file: include.tex
\def\simge{\mathrel{%
   \rlap{\raise 0.511ex \hbox{$>$}}{\lower 0.511ex \hbox{$\sim$}}}}
\def\simle{\mathrel{
   \rlap{\raise 0.511ex \hbox{$<$}}{\lower 0.511ex \hbox{$\sim$}}}}

\newcommand{\pt}{p_T}
\newcommand{\Et}{E_T}

\newcommand {\GeVc} {{\rm GeV}/c}

\newcommand {\GeV} {{\rm GeV}/c^2}

\newcommand{\MET}{\mbox{$E\kern-0.60em\raise0.10ex\hbox{/}_{T}$}}

\newcommand{\W}{W}
\newcommand{\Z}{Z}
\newcommand{\WW}{WW}
\newcommand{\WZ}{WZ}
\newcommand{\ZZ}{ZZ}
\newcommand{\WZdecay}{WZ \rightarrow \ell^{\prime}\nu_{\ell^{\prime}} \ell\ell}

\newcommand{\ppbar}{p \overline{p}}

\newcommand{\ttbar}{t \overline{t}}

\newcommand{\Dzero} {\rm{D}\O}

\newcommand{\etad}{|\eta_{\rm{d}}|}
\newcommand{\Mll}{M_{\ell^{+}\ell^{-}}}

%% file: authorList/authorList.tex
\affiliation{Institute of Physics, Academia Sinica, Taipei, Taiwan 11529, Republic of China} 
\affiliation{Argonne National Laboratory, Argonne, Illinois 60439} 
\affiliation{Institut de Fisica d'Altes Energies, Universitat Autonoma de Barcelona, E-08193, Bellaterra (Barcelona), Spain} 
\affiliation{Baylor University, Waco, Texas  76798} 
\affiliation{Istituto Nazionale di Fisica Nucleare, University of Bologna, I-40127 Bologna, Italy} 
\affiliation{Brandeis University, Waltham, Massachusetts 02254} 
\affiliation{University of California, Davis, Davis, California  95616} 
\affiliation{University of California, Los Angeles, Los Angeles, California  90024} 
\affiliation{University of California, San Diego, La Jolla, California  92093} 
\affiliation{University of California, Santa Barbara, Santa Barbara, California 93106} 
\affiliation{Instituto de Fisica de Cantabria, CSIC-University of Cantabria, 39005 Santander, Spain} 
\affiliation{Carnegie Mellon University, Pittsburgh, PA  15213} 
\affiliation{Enrico Fermi Institute, University of Chicago, Chicago, Illinois 60637} 
\affiliation{Comenius University, 842 48 Bratislava, Slovakia; Institute of Experimental Physics, 040 01 Kosice, Slovakia} 
\affiliation{Joint Institute for Nuclear Research, RU-141980 Dubna, Russia} 
\affiliation{Duke University, Durham, North Carolina  27708} 
\affiliation{Fermi National Accelerator Laboratory, Batavia, Illinois 60510} 
\affiliation{University of Florida, Gainesville, Florida  32611} 
\affiliation{Laboratori Nazionali di Frascati, Istituto Nazionale di Fisica Nucleare, I-00044 Frascati, Italy} 
\affiliation{University of Geneva, CH-1211 Geneva 4, Switzerland} 
\affiliation{Glasgow University, Glasgow G12 8QQ, United Kingdom} 
\affiliation{Harvard University, Cambridge, Massachusetts 02138} 
\affiliation{Division of High Energy Physics, Department of Physics, University of Helsinki and Helsinki Institute of Physics, FIN-00014, Helsinki, Finland} 
\affiliation{University of Illinois, Urbana, Illinois 61801} 
\affiliation{The Johns Hopkins University, Baltimore, Maryland 21218} 
\affiliation{Institut f\"{u}r Experimentelle Kernphysik, Universit\"{a}t Karlsruhe, 76128 Karlsruhe, Germany} 
\affiliation{High Energy Accelerator Research Organization (KEK), Tsukuba, Ibaraki 305, Japan} 
\affiliation{Center for High Energy Physics: Kyungpook National University, Taegu 702-701, Korea; Seoul National University, Seoul 151-742, Korea; and SungKyunKwan University, Suwon 440-746, Korea} 
\affiliation{Ernest Orlando Lawrence Berkeley National Laboratory, Berkeley, California 94720} 
\affiliation{University of Liverpool, Liverpool L69 7ZE, United Kingdom} 
\affiliation{University College London, London WC1E 6BT, United Kingdom} 
\affiliation{Centro de Investigaciones Energeticas Medioambientales y Tecnologicas, E-28040 Madrid, Spain} 
\affiliation{Massachusetts Institute of Technology, Cambridge, Massachusetts  02139} 
\affiliation{Institute of Particle Physics: McGill University, Montr\'{e}al, Canada H3A~2T8; and University of Toronto, Toronto, Canada M5S~1A7} 
\affiliation{University of Michigan, Ann Arbor, Michigan 48109} 
\affiliation{Michigan State University, East Lansing, Michigan  48824} 
\affiliation{Institution for Theoretical and Experimental Physics, ITEP, Moscow 117259, Russia} 
\affiliation{University of New Mexico, Albuquerque, New Mexico 87131} 
\affiliation{Northwestern University, Evanston, Illinois  60208} 
\affiliation{The Ohio State University, Columbus, Ohio  43210} 
\affiliation{Okayama University, Okayama 700-8530, Japan} 
\affiliation{Osaka City University, Osaka 588, Japan} 
\affiliation{University of Oxford, Oxford OX1 3RH, United Kingdom} 
\affiliation{University of Padova, Istituto Nazionale di Fisica Nucleare, Sezione di Padova-Trento, I-35131 Padova, Italy} 
\affiliation{LPNHE, Universite Pierre et Marie Curie/IN2P3-CNRS, UMR7585, Paris, F-75252 France} 
\affiliation{University of Pennsylvania, Philadelphia, Pennsylvania 19104} 
\affiliation{Istituto Nazionale di Fisica Nucleare Pisa, Universities of Pisa, Siena and Scuola Normale Superiore, I-56127 Pisa, Italy} 
\affiliation{University of Pittsburgh, Pittsburgh, Pennsylvania 15260} 
\affiliation{Purdue University, West Lafayette, Indiana 47907} 
\affiliation{University of Rochester, Rochester, New York 14627} 
\affiliation{The Rockefeller University, New York, New York 10021} 
\affiliation{Istituto Nazionale di Fisica Nucleare, Sezione di Roma 1, University of Rome ``La Sapienza," I-00185 Roma, Italy} 
\affiliation{Rutgers University, Piscataway, New Jersey 08855} 
\affiliation{Texas A\&M University, College Station, Texas 77843} 
\affiliation{Istituto Nazionale di Fisica Nucleare, University of Trieste/\ Udine, Italy} 
\affiliation{University of Tsukuba, Tsukuba, Ibaraki 305, Japan} 
\affiliation{Tufts University, Medford, Massachusetts 02155} 
\affiliation{Waseda University, Tokyo 169, Japan} 
\affiliation{Wayne State University, Detroit, Michigan  48201} 
\affiliation{University of Wisconsin, Madison, Wisconsin 53706} 
\affiliation{Yale University, New Haven, Connecticut 06520} 
\author{A.~Abulencia}
\affiliation{University of Illinois, Urbana, Illinois 61801}
\author{J.~Adelman}
\affiliation{Enrico Fermi Institute, University of Chicago, Chicago, Illinois 60637}
\author{T.~Affolder}
\affiliation{University of California, Santa Barbara, Santa Barbara, California 93106}
\author{T.~Akimoto}
\affiliation{University of Tsukuba, Tsukuba, Ibaraki 305, Japan}
\author{M.G.~Albrow}
\affiliation{Fermi National Accelerator Laboratory, Batavia, Illinois 60510}
\author{D.~Ambrose}
\affiliation{Fermi National Accelerator Laboratory, Batavia, Illinois 60510}
\author{S.~Amerio}
\affiliation{University of Padova, Istituto Nazionale di Fisica Nucleare, Sezione di Padova-Trento, I-35131 Padova, Italy}
\author{D.~Amidei}
\affiliation{University of Michigan, Ann Arbor, Michigan 48109}
\author{A.~Anastassov}
\affiliation{Rutgers University, Piscataway, New Jersey 08855}
\author{K.~Anikeev}
\affiliation{Fermi National Accelerator Laboratory, Batavia, Illinois 60510}
% ...
\collaboration{CDF Collaboration\footnote{With visitors from $^a$University of Athens, 
$^b$University of Bristol, 
$^c$University Libre de Bruxelles, 
$^d$Cornell University, 
$^e$University of Cyprus, 
$^f$University of Dublin, 
$^g$University of Edinburgh, 
$^h$University of Heidelberg, 
$^i$Universidad Iberoamericana, 
$^j$University of Manchester, 
$^k$Nagasaki Institute of Applied Science, 
$^l$University de Oviedo, 
$^m$University of London, Queen Mary and Westfield College, 
$^n$Texas Tech University, 
$^o$IFIC(CSIC-Universitat de Valencia), 
}}
\noaffiliation

%% file: WZ_control_regions.tex
\begin{table}
\caption{Summary of the expected and observed yields in the trilepton control
regions. In the classification column, $\ell_t$ denotes a track-only lepton
candidate having unknown flavor. The $eee$, $e\mu\mu$, $ee\ell_t$, and
$e\mu \ell_t$ classifications receive a large contribution from $Z\gamma$
events where the photon is reconstructed as a forward electron without
a matched track.
\label{tbl:control_regions}}
\begin{ruledtabular}
\begin{tabular}{lrrrr}
Flavor & \multicolumn{2}{c}{$Z$-mass} &
\multicolumn{2}{c}{$Z$-veto}  \\
Classification          &  Expected                  &  Data    &  Expected          &  Data         \\
\hline\hline
$e$ $e$ $e$       &   116.3  $\pm$  19.2    &  103     &   114.8  $\pm$  22.5    &  103  \\
$e$ $e$ $\mu$     &     1.8  $\pm$   0.3    &   2      &     1.4  $\pm$   0.4    &   4  \\
$e$ $\mu$ $\mu$   &    62.5  $\pm$  10.3    &  50      &    69.2  $\pm$  14.0    &  62  \\
$\mu$ $\mu$ $\mu$ &     1.1  $\pm$   0.2    &   1      &     0.3  $\pm$   0.1    &   1  \\
$e$ $e$ $\ell_t$       &    29.6  $\pm$   4.6    &  20      &    33.5  $\pm$   6.2    &  31  \\
$e$ $\mu$ $\ell_t$     &    24.9  $\pm$   4.1    &  33      &    26.5  $\pm$   5.2    &  34  \\
$\mu$ $\mu$ $\ell_t$   &     2.7  $\pm$   0.4    &   5      &     1.9  $\pm$   0.4    &   3  \\
$e$ $\ell_t$ $\ell_t$       &     4.0  $\pm$   0.7    &   1      &     2.6  $\pm$   0.5    &   2  \\
$\mu$ $\ell_t$ $\ell_t$     &     0.4  $\pm$   0.2    &   0      &     0.4  $\pm$   0.1    &   1  \\
\hline
Total             &   243.5  $\pm$  38.8    &  215     &   250.9  $\pm$  48.3    &  241  \\
\end{tabular}
\end{ruledtabular}
\end{table}

%% file: WZ_expect_withsys.tex
\begin{table}
\caption{Expected number of events in the signal region for $WZ$ and
the background contributions. ``Lumi'' refers to the integrated
luminosity uncertainty, which is absent for the $Z$+jets because it is
determined from the same dataset.
\label{tbl:results}}
\begin{ruledtabular}
\begin{tabular}{lc}
  Source               &  Expectation $\pm$ Stat $\pm$ Syst $\pm$ Lumi \\
\hline\hline
  $Z$+jets             &  1.21 $\pm$ 0.27  $\pm$ 0.28  $\pm$ ~$-$~ \\
  $ZZ$                 &  0.88 $\pm$ 0.01  $\pm$ 0.09  $\pm$ 0.05 \\
  $Z\gamma$            &  0.44 $\pm$ 0.05  $\pm$ 0.15  $\pm$ 0.03 \\
  $t\bar{t}$           &  0.12 $\pm$ 0.01  $\pm$ 0.02  $\pm$ 0.01 \\
\hline
  Total Background     &  2.65 $\pm$ 0.28  $\pm$ 0.33  $\pm$ 0.09  \\
\hline
  $WZ$                 &  9.75 $\pm$ 0.03  $\pm$ 0.31  $\pm$ 0.59 \\
\hline
  Total Expected       &  12.41 $\pm$ 0.28  $\pm$ 0.45  $\pm$ 0.67  \\
\hline
Observed               &   16 \\
\end{tabular}
\end{ruledtabular}
\end{table}

%% file: document.bbl
\begin{thebibliography}{19}
\expandafter\ifx\csname natexlab\endcsname\relax\def\natexlab#1{#1}\fi
\expandafter\ifx\csname bibnamefont\endcsname\relax
  \def\bibnamefont#1{#1}\fi
\expandafter\ifx\csname bibfnamefont\endcsname\relax
  \def\bibfnamefont#1{#1}\fi
\expandafter\ifx\csname citenamefont\endcsname\relax
  \def\citenamefont#1{#1}\fi
\expandafter\ifx\csname url\endcsname\relax
  \def\url#1{\texttt{#1}}\fi
\expandafter\ifx\csname urlprefix\endcsname\relax\def\urlprefix{URL }\fi
\providecommand{\bibinfo}[2]{#2}
\providecommand{\eprint}[2][]{\url{#2}}

\bibitem[{\citenamefont{Acosta~{\it et
  al.}}(2005{\natexlab{a}})}]{Acosta:2004uq}
\bibinfo{author}{\bibfnamefont{D.}~\bibnamefont{Acosta~{\it et al.}}}
  (\bibinfo{collaboration}{CDF Collaboration}), \bibinfo{journal}{Phys. Rev.
  Lett.} \textbf{\bibinfo{volume}{94}}, \bibinfo{pages}{091803}
  (\bibinfo{year}{2005}{\natexlab{a}}).

\bibitem[{\citenamefont{Hagiwara et~al.}(1993)\citenamefont{Hagiwara, Ishihara,
  Szalapski, and Zeppenfeld}}]{PhysRevD.48.2182}
\bibinfo{author}{\bibfnamefont{K.}~\bibnamefont{Hagiwara}},
  \bibinfo{author}{\bibfnamefont{S.}~\bibnamefont{Ishihara}},
  \bibinfo{author}{\bibfnamefont{R.}~\bibnamefont{Szalapski}},
  \bibnamefont{and}
  \bibinfo{author}{\bibfnamefont{D.}~\bibnamefont{Zeppenfeld}},
  \bibinfo{journal}{Phys. Rev. D} \textbf{\bibinfo{volume}{48}},
  \bibinfo{pages}{2182} (\bibinfo{year}{1993}).

\bibitem[{\citenamefont{Abazov~{\it et al.}}(2005)}]{Abazov:2005ys}
\bibinfo{author}{\bibfnamefont{V.~M.} \bibnamefont{Abazov~{\it et al.}}}
  (\bibinfo{collaboration}{D\O\ Collaboration}), \bibinfo{journal}{Phys. Rev.
  Lett.} \textbf{\bibinfo{volume}{95}}, \bibinfo{pages}{141802}
  (\bibinfo{year}{2005}).

\bibitem[{\citenamefont{Acosta~{\it et
  al.}}(2005{\natexlab{b}})}]{Acosta:2005pq}
\bibinfo{author}{\bibfnamefont{D.}~\bibnamefont{Acosta~{\it et al.}}}
  (\bibinfo{collaboration}{CDF Collaboration}), \bibinfo{journal}{Phys. Rev. D}
  \textbf{\bibinfo{volume}{71}}, \bibinfo{pages}{091105}
  (\bibinfo{year}{2005}{\natexlab{b}}).

\bibitem[{\citenamefont{Campbell and Ellis}(1999)}]{Campbell:1999ah}
\bibinfo{author}{\bibfnamefont{J.~M.} \bibnamefont{Campbell}} \bibnamefont{and}
  \bibinfo{author}{\bibfnamefont{R.~K.} \bibnamefont{Ellis}},
  \bibinfo{journal}{Phys. Rev. D} \textbf{\bibinfo{volume}{60}},
  \bibinfo{pages}{113006} (\bibinfo{year}{1999}).

\bibitem[{\citenamefont{Acosta~{\it et
  al.}}(2005{\natexlab{c}})}]{Acosta:2004yw}
\bibinfo{author}{\bibfnamefont{D.}~\bibnamefont{Acosta~{\it et al.}}}
  (\bibinfo{collaboration}{CDF Collaboration}), \bibinfo{journal}{Phys. Rev. D}
  \textbf{\bibinfo{volume}{71}}, \bibinfo{pages}{032001}
  (\bibinfo{year}{2005}{\natexlab{c}}).

\bibitem[{\citenamefont{Sill~{\it et al.}}(2000)}]{SVXII}
\bibinfo{author}{\bibfnamefont{A.}~\bibnamefont{Sill~{\it et al.}}},
  \bibinfo{journal}{Nucl. Instrum. Methods A} \textbf{\bibinfo{volume}{447}},
  \bibinfo{pages}{1} (\bibinfo{year}{2000}).

\bibitem[{\citenamefont{Affolder~{\it et al.}}(2000)}]{ISL}
\bibinfo{author}{\bibfnamefont{A.}~\bibnamefont{Affolder~{\it et al.}}},
  \bibinfo{journal}{Nucl. Instrum. Methods A} \textbf{\bibinfo{volume}{453}},
  \bibinfo{pages}{84} (\bibinfo{year}{2000}).

\bibitem[{\citenamefont{Affolder~{\it et al.}}(2004)}]{COT}
\bibinfo{author}{\bibfnamefont{T.}~\bibnamefont{Affolder~{\it et al.}}},
  \bibinfo{journal}{Nucl. Instrum. Methods A} \textbf{\bibinfo{volume}{526}},
  \bibinfo{pages}{249} (\bibinfo{year}{2004}).

\bibitem[{\citenamefont{Yao~{\it et al.}}(2006)}]{PDG}
\bibinfo{author}{\bibfnamefont{W.-M.} \bibnamefont{Yao~{\it et al.}}}
  (\bibinfo{collaboration}{Particle Data Group}), \bibinfo{journal}{J. Phys. G}
  \textbf{\bibinfo{volume}{33}}, \bibinfo{pages}{1} (\bibinfo{year}{2006}).

\bibitem[{\citenamefont{Brun et~al.}()\citenamefont{Brun, Hagelberg, Hansroul,
  and Lassalle}}]{GEANT}
\bibinfo{author}{\bibfnamefont{R.}~\bibnamefont{Brun}},
  \bibinfo{author}{\bibfnamefont{R.}~\bibnamefont{Hagelberg}},
  \bibinfo{author}{\bibfnamefont{M.}~\bibnamefont{Hansroul}}, \bibnamefont{and}
  \bibinfo{author}{\bibfnamefont{J.}~\bibnamefont{Lassalle}}, \eprint{version
  3.15, CERN-DD-78-2-REV}.

\bibitem[{\citenamefont{Sjostrand et~al.}(2006)\citenamefont{Sjostrand, Mrenna,
  and Skands}}]{Sjostrand:2006za}
\bibinfo{author}{\bibfnamefont{T.}~\bibnamefont{Sjostrand}},
  \bibinfo{author}{\bibfnamefont{S.}~\bibnamefont{Mrenna}}, \bibnamefont{and}
  \bibinfo{author}{\bibfnamefont{P.}~\bibnamefont{Skands}},
  \bibinfo{journal}{J. High Energy Phys.} \textbf{\bibinfo{volume}{0605}},
  \bibinfo{pages}{026} (\bibinfo{year}{2006}).

\bibitem[{\citenamefont{Baur and Berger}(1993)}]{Baur:1992cd}
\bibinfo{author}{\bibfnamefont{U.}~\bibnamefont{Baur}} \bibnamefont{and}
  \bibinfo{author}{\bibfnamefont{E.~L.} \bibnamefont{Berger}},
  \bibinfo{journal}{Phys. Rev. D} \textbf{\bibinfo{volume}{47}},
  \bibinfo{pages}{4889} (\bibinfo{year}{1993}).

\bibitem[{\citenamefont{Lai~{\it et al.}}(2000)}]{Lai:1999wy}
\bibinfo{author}{\bibfnamefont{H.~L.} \bibnamefont{Lai~{\it et al.}}}
  (\bibinfo{collaboration}{CTEQ Collaboration}), \bibinfo{journal}{Eur. Phys.
  J. C} \textbf{\bibinfo{volume}{12}}, \bibinfo{pages}{375}
  (\bibinfo{year}{2000}).

\bibitem[{\citenamefont{Baur et~al.}(1998)\citenamefont{Baur, Han, and
  Ohnemus}}]{Baur:1997kz}
\bibinfo{author}{\bibfnamefont{U.}~\bibnamefont{Baur}},
  \bibinfo{author}{\bibfnamefont{T.}~\bibnamefont{Han}}, \bibnamefont{and}
  \bibinfo{author}{\bibfnamefont{J.}~\bibnamefont{Ohnemus}},
  \bibinfo{journal}{Phys. Rev. D} \textbf{\bibinfo{volume}{57}},
  \bibinfo{pages}{2823} (\bibinfo{year}{1998}).

\bibitem[{\citenamefont{Kidonakis and Vogt}(2003)}]{Kidonakis:2003qe}
\bibinfo{author}{\bibfnamefont{N.}~\bibnamefont{Kidonakis}} \bibnamefont{and}
  \bibinfo{author}{\bibfnamefont{R.}~\bibnamefont{Vogt}},
  \bibinfo{journal}{Phys. Rev. D} \textbf{\bibinfo{volume}{68}},
  \bibinfo{pages}{114014} (\bibinfo{year}{2003}).

\bibitem[{\citenamefont{Cacciari et~al.}(2004)\citenamefont{Cacciari, Frixione,
  Mangano, Nason, and Ridolfi}}]{Cacciari:2003fi}
\bibinfo{author}{\bibfnamefont{M.}~\bibnamefont{Cacciari}},
  \bibinfo{author}{\bibfnamefont{S.}~\bibnamefont{Frixione}},
  \bibinfo{author}{\bibfnamefont{M.~L.} \bibnamefont{Mangano}},
  \bibinfo{author}{\bibfnamefont{P.}~\bibnamefont{Nason}}, \bibnamefont{and}
  \bibinfo{author}{\bibfnamefont{G.}~\bibnamefont{Ridolfi}},
  \bibinfo{journal}{J. High Energy Phys.} \textbf{\bibinfo{volume}{0404}},
  \bibinfo{pages}{068} (\bibinfo{year}{2004}).

\bibitem[{\citenamefont{Kretzer et~al.}(2004)\citenamefont{Kretzer, Lai,
  Olness, and Tung}}]{Kretzer:2003it}
\bibinfo{author}{\bibfnamefont{S.}~\bibnamefont{Kretzer}},
  \bibinfo{author}{\bibfnamefont{H.~L.} \bibnamefont{Lai}},
  \bibinfo{author}{\bibfnamefont{F.~I.} \bibnamefont{Olness}},
  \bibnamefont{and} \bibinfo{author}{\bibfnamefont{W.~K.} \bibnamefont{Tung}}
  (\bibinfo{collaboration}{CTEQ}), \bibinfo{journal}{Phys. Rev. D}
  \textbf{\bibinfo{volume}{69}}, \bibinfo{pages}{114005}
  (\bibinfo{year}{2004}).

\bibitem[{\citenamefont{Acosta~{\it et al.}}(2002)}]{Acosta:2002hx}
\bibinfo{author}{\bibfnamefont{D.}~\bibnamefont{Acosta~{\it et al.}}},
  \bibinfo{journal}{Nucl. Instrum. Methods A} \textbf{\bibinfo{volume}{494}},
  \bibinfo{pages}{57} (\bibinfo{year}{2002}).

\end{thebibliography}
